\documentclass[12pt]{article}

\newcommand\nice[1]{#1}    \newcommand\subm[1]{}   
\subm{ \renewcommand\dosingle[1]{ }    }

\usepackage{graphicx}

\usepackage{amsfonts}
\usepackage{mathpi}

\usepackage[numbers]{natbib}  
\usepackage{url} 

\usepackage{hyperref}

\usepackage[symbol]{footmisc}

\providecommand{\eprint}[1]{\href{http://arxiv.org/abs/#1}{[arXiv:#1]}}

\providecommand{\url}[1]{\href{#1}{#1}}

\providecommand{\adsurl}[1]{} 

\newcommand\BIBAABST{aa_hyperref} 
\newcommand\BIBHOME{mybib} 

\def\apjs{ApJSupp}                 
\def\aap{A\&A}            
\def\mnras{MNRAS}

\def\prd{Phys.Rev.D}
\def\physrep{Phys.Reports}
\def\nat{Nature}

\nice{  }

\newcommand\gtapprox{\,\lower.6ex\hbox{$\buildrel >\over \sim$} \, }
\newcommand\ltapprox{\,\lower.6ex\hbox{$\buildrel <\over \sim$} \, }
\newcommand\propapprox{\,\lower.6ex\hbox{$\buildrel \propto\over \sim$} \, }

\newcommand\arcs{\ifmmode {'' }\else $'' $\fi}     
\newcommand\arcm{\ifmmode {' }\else $' $\fi}       
\newcommand\ddeg{\ifmmode^\circ\else$^\circ$\fi}    

\newcommand\frtoday{Le\space\number\day\space\ifcase\month\or
  janvier\or f\'evrier\or mars\or avril\or mai\or juin\or
  juillet\or ao\^ut\or septembre\or octobre\or novembre\or 
d\'ecembre\fi\space \number\year}

\newcommand\cqg{ClassQuantGra}   %
\newcommand\hGpc{\mbox{$h^{-1}$ Gpc}}

\newcommand\Omtot{\Omega_{\mbox{\rm \small tot}}}

\title{Does gravity prefer the Poincar\'e dodecahedral space?}

\author{Boudewijn F. Roukema \\
{\em Toru\'n Centre for Astronomy, Nicolaus Copernicus University}, \\
{\em ul. Gagarina 11, 87-100 Toru\'n, Poland} \\
{\tt boud@astro.uni.torun.pl}\\
28 March 2009
}

\date{{\it Essay written for the Gravity Research Foundation}\\
{\it 2009 Awards for Essays on Gravitation;} \\
{\it awarded an Honorable Mention}
}

\begin{document}

\maketitle

\abstract{The missing fluctuations problem in cosmic microwave
  background observations is naturally explained by well-proportioned
  small universe models. Among the well-proportioned models, the
  Poincar\'e dodecahedral space is empirically favoured. Does gravity
  favour this space? The residual gravity effect is the residual
  acceleration induced by weak limit gravity from multiple topological
  images of a massive object on a nearby negligible mass test object.
  At the present epoch, the residual gravity effect is about a million
  times weaker in three of the well-proportioned spaces than in
  ill-proportioned spaces. However, in the Poincar\'e space, the effect is
  10,000 times weaker still, i.e. the Poincar\'e space is about
  $10^{10}$ times ``better balanced'' than ill-proportioned
  spaces. Both observations and weak limit dynamics select the
  Poincar\'e space to be special.}


\clearpage


\section{The missing fluctuations problem and well-proportioned spaces}

Through the Einstein field equations, differential geometry and
astronomical observations have converged during the past decade on
the concordance model of physical cosmology
\cite{CosConcord95}. The concordance model matches an impressive range
of astronomical observational data sets, including both the cosmic microwave
background (e.g., \cite{WMAPSpergel}) and surveys of gravitationally
collapsed astrophysical objects.
Nevertheless, 
the concordance model is seriously
incomplete: it does not say what 3-manifold
describes the comoving space that we inhabit. At best, it only chooses
between the three classes of constant curvature 3-manifolds,
i.e. between those of negative, zero and positive curvature. The barely
noticed ``elephant in the room'' is the topology of comoving space.
This question was raised by Karl Schwarzschild
\cite{Schw00,Schw98}, but has been studied mostly 
during the last decade and a half
(e.g. \cite{LaLu95,Lum98,Stark98,LR99,BR99,RG04}).
What are comoving space's global symmetries -- i.e. the holonomy
transformations, which are isometries that map objects to themselves
in the simply connected covering space, which is the apparent space
from the ``na\"{\i}ve'' observer's point of view (something like
the apparent space seen in a mirror-lined room)? 

A century ago, the missing ether problem revealed by
the Michelson-Morley experiment was solved by dropping 
the assumption that space and time are independent.
Now we have the missing fluctuations problem (e.g. \cite{Copi07,Copi09}, 
and references therein) of the 
COsmic Background Explorer (COBE) and Wilkinson Microwave Anisotropy Probe (WMAP)
all-sky cosmic microwave background (CMB) experiments. 
Cosmic topology solves this problem. It also leads to dropping the assumption
that dynamics is independent of the global topology of comoving space.

The missing fluctuations problem is solved by the generic prediction 
that in a multiply connected 3-manifold,
correlations between density perturbations should vanish above a
length scale similar to that of the size of the
3-manifold,\footnote{See fig.~10 of \protect\cite{LR99} for various
  definitions of ``size''.} since {\em (comoving) objects larger than 
space itself cannot exist}. 
This argument is weakened, but remains approximately valid in the observer's apparent space,
in particular on the surface of last scattering (SLS),
which is a thin 2-spherical shell of radius about 10{\hGpc}
(comoving) centred at the observer.  
CMB fluctuations are seen primarily on the (SLS).

A cut-off in correlations at
large scales was suspected in the COBE data, and confirmed in the WMAP
data (Fig.~16, \cite{WMAPSpergel}). Estimates of the chance
of this lack of large angular scale (``low $l$'') structure occurring
in an infinite, flat model range from $0.3\%$ (Sect. 7,
\cite{WMAPSpergel}) to $12.5\%$ (Table 5, \cite{EfstNoProb03b}) for
the first-year WMAP data, decreasing to $0.03\%$ for the three-year
and five-year
WMAP data \cite{Copi07,Copi09}.  


Not all multiply connected spaces with a short length give a strong
cut-off effect. Spaces whose fundamental lengths are approximately 
equal, called ``well-proportioned'', are the most likely to provide 
a large-scale cut-off in structure statistics \cite{WeeksWellProp04}.

\section{The Poincar\'e dodecahedral space $S^3/I^*$}

Estimates of the curvature of the
Universe on the scale of the SLS, via the total density parameter
$\Omtot$, hint at positive curvature, e.g. $\Omtot =
1.014\pm0.017$ from the WMAP three-year data together with Hubble
Space Telescope key project estimates of the Hubble constant $H_0$
(Table 12, \cite{WMAPSpergel06});
$0.9915 < \Omtot < 1.0175$
from combining the WMAP five-year data, 
baryonic acoustic oscillations in galaxy surveys and supernovae
data \citep{WMAP5Komatsu}. The missing fluctuations problem 
and the curvature estimates 
led to the proposal of one of the well-proportioned spaces, 
the Poincar\'e Dodecahedral Space (PDS),
$S^3 /I^*$,\footnote{$I^*$ is the binary icosahedral group.} \citep{LumNat03}.
as a candidate for comoving space. 

The PDS has a positively curved solid dodecahedron as its fundamental
domain. Several groups have investigated this model 
\cite{RLCMB04,Aurich2005a,Aurich2005b,Gundermann2005,KeyCSS06,Caillerie07,NJ07,LR08}.
When thinking of the PDS fundamental domain projected to $\mathbb{R}^3$,
the identification of opposite faces must take place by a translation
followed by a rotation of $\pm \pi/5 = \pm 36\ddeg$. If the model is correct, then
despite the observed {\em average} lack of correlations on large scales, 
the correlations in {\em certain directions} should be high,
since certain regions of comoving space are multiply viewed.

The exact set of points seen twice is defined by the identified
circles principle \cite{Corn96,Corn98b}, but a larger amount of
information in the WMAP data can be used by 
cross-correlating temperature fluctuations between adjacent copies
of the SLS \cite{RBSG08,RBG08} in apparent space.  This method gives
an optimal astronomical orientation of the fundamental dodecahedron,
and by allowing the search algorithm (Markov chain Monte Carlo) to
investigate arbitrary twists (i.e. not constrained to $\pm 36\ddeg$),
it yields an optimal twist when matching opposite faces.  The optimal
orientation found in the WMAP data for the fundamental domain gives a strong
cross-correlation, i.e.  strong correlations exist
between apparently distant points on the sky 
in a small number of directions, despite the missing fluctuations problem,
and the optimal twist angle is $(+39\pm
2.5)\ddeg$, consistent with that required, despite the freedom allowed
by the search algorithm \cite{RBSG08,RBG08}.

Is this just an empirically preferred space, or could the Poincar\'e
space be favoured by gravity?


\section{The residual gravity acceleration effect}

It has been shown heuristically that global topology in a universe
containing at least one density perturbation can feed back on 
local dynamics \cite{RBBSJ06}. This can be seen most easily 
in a $T^1 \times \mathbb{R}^2$ model of length $L$, considering a massive object 
of mass $m$,
its two adjacent images in the covering space $\mathbb{R}^3$, 
and a massless test particle displaced $x$ from the massive object
in the short direction (Fig.~3, \cite{RBBSJ06}. 
In the (Newtonian) weak limit, in addition to 
being accelerated by the ``local'' copy of the massive object, the 
test particle is pulled in opposite directions towards the two 
distant copies of the massive particle. The latter two accelerations
are nearly, but not quite, equal. {\em The net effect is that the 
test particle has a small extra pull towards the closer of the two 
distant images of the massive object, i.e. it falls towards the 
``local'' copy of the massive
object more slowly than would be expected if multiple images were ignored.}
This is the ``residual gravity effect''.  In $T^1 \times \mathbb{R}^2$, 
for an object of fixed mass $m$, to first order in $x/L$,
where $x/L \ll 1$,
the residual acceleration $\ddot{x}$ is proportional to $x/L$. 

What happens in other spaces, in particular the well-proportioned spaces?
%
%
%
Perfectly regular well-proportioned \citep{WeeksWellProp04}
spaces $T^3$, 
$S^3/T^*$,
$S^3/O^*$, and
$S^3/I^*$ are ``better balanced'' than ill-proportioned spaces
such as $T^1\times \mathbb{R}^2$.
When considering all the adjacent topological images
of a massive object and a negligible mass test particle displaced from it 
slightly in an 
arbitrary direction \cite{RBBSJ06,RR09}, 
the first order term
in $x/L$ (for $T^3$) or $x/r_C$ (for the spherical spaces,
with curvature radius $r_C$) of the residual gravity effect vanishes.
Small perturbations from 
perfect isotropy destroy this equilibrium. However, they induce
a first order effect that tends to oppose the anisotropy and 
restore the equilibrium, favouring an equilibrium state in which the
first order term cancels to zero.

However, what is especially surprising is that one of these four
well-proportioned spaces is ``more equal than the others''. The
highest order term for the residual gravity effect in $T^3$,
$S^3/T^*$, and $S^3/O^*$ is the third order term, but in the
Poincar\'e space $S^3/I^*$, the third order term cancels, leaving
the fifth order as the highest term \citep{RR09}. Hence, at the
present epoch,\footnote{For a displacement relative 
to the cosmic web and observable cosmic topology, 
$x/L \sim x/r_C \sim 10^{-3}$.}
not only is the residual gravity effect about a million
times weaker in three of the well-proportioned spaces than in
ill-proportioned spaces, but in the Poincar\'e space, the effect is
10,000 times weaker still, i.e. {\em the Poincar\'e space is at present about
  $10^{10}$ times ``better balanced'' than ill-proportioned spaces}.

\section{Conclusion}

Through the residual gravity effect, global topology can affect dynamics.
Moreover, the effect singles out a special role for the Poincar\'e
space.
It is likely that the effect was most relevant during
early, pre-inflationary epochs. If weak limit gravity were physically relevant
and if inhomogeneities existed at those epochs, then dynamics could have
selected the Poincar\'e space as the best balanced 3-manifold, especially during
the quantum epoch.  The Poincar\'e space
is also the space that seems to be favoured by observations. Is this
just a coincidence, or are the missing fluctuations above scales of 
60{\ddeg} in the WMAP data a sign that gravity selected (the comoving spatial
section of) the Universe to be a Poincar\'e dodecahedral space?


\bibliographystyle{\BIBAABST} 
\bibliography{\BIBHOME}


\end{document}